\documentclass[aps,twocolumn,showpacs,byrevtex]{revtex4-1}
\usepackage[colorlinks,linkcolor=blue,anchorcolor=blue,citecolor=blue]{hyperref}
\usepackage{lineno}
\usepackage{times}
\usepackage{amsmath}
\usepackage{graphicx}
\usepackage{float}
\usepackage{color}

\usepackage[T1]{fontenc}

\newcommand{\x}{\chi_{c1}(3872)}
\newcommand{\pp}{\pi^+\pi^-}
\newcommand{\pip}{\pi^+}
\newcommand{\pim}{\pi^-}

\newcommand{\LL}{\ell^+\ell^-}
\newcommand{\EE}{e^+e^-}
\newcommand{\ee}{e^+e^-}

\newcommand{\mm}{\mu^+\mu^-}

\newcommand{\psip}{\psi(2S)}

\newcommand{\jpsi}{J/\psi}
\newcommand{\piz}{\pi^0}
\newcommand{\chico}{\chi_{c1}}

\newcommand{\ppjpsi}{\pi^+\pi^-J/\psi}

\def\Journal#1#2#3#4{{#1} {\bf #2}, #3 (#4)}

\def\PRD{Phys. Rev. D}

\begin{document}

\title{\boldmath Search for the decay $\chi_{c1}(3872)\to\pi^{+}\pi^{-}\chi_{c1}$
}

    \author{M.~Ablikim$^{1}$, M.~N.~Achasov$^{4,b}$, P.~Adlarson$^{75}$, X.~C.~Ai$^{81}$, R.~Aliberti$^{35}$, A.~Amoroso$^{74A,74C}$, M.~R.~An$^{39}$, Q.~An$^{71,58}$, Y.~Bai$^{57}$, O.~Bakina$^{36}$
% ...
    	\\
    	\vspace{0.2cm}
    	(BESIII Collaboration)\\
    	\vspace{0.2cm} {\it
    		$^{1}$ Institute of High Energy Physics, Beijing 100049, People's Republic of China\\
% ...
$^{4}$ Budker Institute of Nuclear Physics SB RAS (BINP), Novosibirsk 630090, Russia\\
% ...
$^{35}$ Johannes Gutenberg University of Mainz, Johann-Joachim-Becher-Weg 45, D-55099 Mainz, Germany\\
    		$^{36}$ Joint Institute for Nuclear Research, 141980 Dubna, Moscow region, Russia\\
% ...
$^{39}$ Liaoning Normal University, Dalian 116029, People's Republic of China\\
% ...
$^{57}$ Southeast University, Nanjing 211100, People's Republic of China\\
    		$^{58}$ State Key Laboratory of Particle Detection and Electronics, Beijing 100049, Hefei 230026, People's Republic of China\\
% ...
$^{71}$ University of Science and Technology of China, Hefei 230026, People's Republic of China\\
% ...
$^{74}$ University of Turin and INFN, (A)University of Turin, I-10125, Turin, Italy; (B)University of Eastern Piedmont, I-15121, Alessandria, Italy; (C)INFN, I-10125, Turin, Italy\\
    		$^{75}$ Uppsala University, Box 516, SE-75120 Uppsala, Sweden\\
% ...
$^{81}$ Zhengzhou University, Zhengzhou 450001, People's Republic of China\\    		
    		\vspace{0.2cm}
% ...
$^{b}$ Also at the Novosibirsk State University, Novosibirsk, 630090, Russia\\
% ...
    	}
        }

\date{\today}

\begin{abstract}

Using a data sample corresponding to an integrated luminosity of 10.9 fb$^{-1}$ 
collected at center-of-mass energies from 4.16 to 4.34 GeV with the BESIII detector, we search for the decay 
$\chi_{c1}(3872) \to \pi^{+}\pi^{-}\chi_{c1}$ in the radiative production $e^{+}e^{-} \to \gamma \chi_{c1}(3872)$. 
No significant signal is observed, and the ratio for the branching fraction of 
$\chi_{c1}(3872) \to \pi^{+}\pi^{-}\chi_{c1}$ to $\chi_{c1}(3872) \to \pi^{+}\pi^{-}J/\psi$ is measured as 
$\mathcal{R}\equiv\frac{\mathcal{B}[\chi_{c1}(3872) \to \pi^{+}\pi^{-}\chi_{c1}]}{\mathcal{B}[\chi_{c1}(3872) \to \pi^{+}\pi^{-} J/\psi]}<0.18$ 
at 90$\%$ confidence level.
An upper limit on the product of the cross section $\sigma[e^{+}e^{-}\to\gamma\chi_{c1}(3872)]$ and the branching fraction 
$\mathcal{B}[\chi_{c1}(3872)\to\pi^{+}\pi^{-}\chi_{c1}]$ at each center-of-mass energy is also given. 
These measurements favor the non-conventional charmonium nature of the $\chi_{c1}(3872)$ state.
\end{abstract}


\maketitle
The $\x$ charmonium-like state was first discovered in 2003 by the Belle experiment 
in the decay $B\to K\x \to K \pip\pim J/\psi$~\cite{Belle-x3872}. 
It was then observed by many experiments in various production and decay modes~\cite{pdg, omegaX}.
The main features of the $\x$  are the following:
it is a narrow state, $\Gamma=1.19 \pm 0.21$~MeV; 
its mass is very close to the 
$D^0\bar{D}^{*0}$ threshold, with an error of 0.18~MeV/$c^2$;
and it has quantum numbers $J^{PC} = 1^{++}$.  
Also, there is an obvious isospin-violation effect in its decays, e.g.~the decay $\x \to \pip\pim J/\psi$ has been found to proceed predominantly via $\rho J/\psi$~\cite{pdg}. 
The experimentally well-established decay channels of the $\x$ 
include $\x\to\pp J/\psi$~\cite{Belle-x3872, BaBar-LHC-ppjpsi, gammax3872}, $ D^0\bar{D}^{*0}$~\cite{BaBar-BELLe-DDbar, BESIII-DDbar-gamJpsi}, $\gamma J/\psi$~\cite{BESIII-DDbar-gamJpsi, BaBar-Belle-LHCb-gamJpsi}, $\pi^0 \chico$~\cite{BESIII-pi0chic1}, and $\omega J/\psi$~\cite{BaBar-wjpsi, BESIII-wjpsi}. 
Theoretical interpretations of the nature of the $\x$ state include a tetraquark state~\cite{theo-4quark}, the $\chi_{c1}(2P)$ charmonium state~\cite{theo-charmo},  and a $D^{0}\bar{D}^{*0}$ molecule state~\cite{theo-mole, theo-mole-2}, without a definitive conclusion. 
Therefore, more detailed studies of $\x$ from both experimental and theoretical sides are necessary for an in-depth 
knowledge of the $\x$ internal structure. 

Unlike $\x \to \omega J/\psi$, the decay $\x \to \piz \chi_{c1}$ is
isospin-violating and presumed to be highly suppressed. 
However, the BESIII experiment has observed the decay $\x \to \piz\chi_{c1}$, 
with a relative branching fraction ratio
$\frac{\mathcal{B}[\x \to \piz \chi_{c1}]}{\mathcal{B}[\x \to \pip\pim J/\psi]}$ = $0.88^{+0.33}_{-0.27}\pm 0.10$~\cite{BESIII-pi0chic1}. Taking into account  $\frac{\mathcal{B}[\x\to\omega\jpsi]}{\mathcal{B}[\x\to\ppjpsi]}= 1.6^{+0.4}_{-0.3}\pm 0.2$~\cite{BESIII-wjpsi}, 
the branching fraction $\mathcal{B}[\x \to \piz \chi_{c1}]$ has a size similar 
to $\mathcal{B}[\x\to\omega\jpsi]$. This unexpectedly large isospin-violation 
indicates that $\x$ may not be a conventional charmonium.  

If the $\x$ is the  $\chico(2P)$ charmonium state, isospin-conserving 
decays, such as $\x \to \pip\pim\chi_{c1}$, are expected to dominate. 
Theoretical calculations show that the one-pion transition 
is significantly suppressed with respect to the di-pion transition, with
$\frac{\Gamma(2^{3}P_{1} \to \chi_{c1}\piz)}{\Gamma(2^{3}P_{1} \to \chi_{c1}\pip\pim)} \approx 0.04$~\cite{pi0/pipi}. 
However, the one-pion transition rate could be enhanced assuming the $\x$ is a shallow
bound state of a $\bar{D^0}D^{*0}$ pair~\cite{X3872-mole}.  Therefore experimental studies of the decay $\x \to \pip\pim\chi_{c1}$ help to discriminate 
theoretical interpretations for the nature of the $\x$.

In this article, we report on the search for the process $\EE\to\gamma\x\to \gamma(\pp\chico)$.
The $\chico$ candidate is reconstructed via its decay to $\gamma\jpsi$, with $\jpsi\to \LL$ ($\ell=e$ or $\mu$).
Due to the relatively low momenta, pions may not be fully detected.
To increase the signal yield, events with either one or two detected pions are used for signal reconstruction.  
The data sample is taken with the BESIII detector~\cite{bes3-detector} at fifteen  center-of-mass (c.m.) energies ranging from $\sqrt{s}=4.16$ to $4.34$~GeV~\cite{ecm},
corresponding to an integrated luminosity of 10.9 $\rm fb^{-1}$~\cite{lum}.

The BESIII detector~\cite{bes3-detector} records symmetric $e^+e^-$ collisions 
provided by the BEPCII storage ring~\cite{Yu:IPAC2016-TUYA01}
in the c.m.~energy region from 2.0 to 4.95~GeV,
with a peak luminosity of $1 \times 10^{33}\;\text{cm}^{-2}\text{s}^{-1}$ 
achieved at $\sqrt{s} = 3.773\;\text{GeV}$. 
 The cylindrical core of the BESIII detector covers 93\% of the full solid angle and consists of a helium-based
multilayer drift chamber~(MDC), a plastic scintillator time-of-flight
system~(TOF), and a CsI(Tl) electromagnetic calorimeter~(EMC),
which are all enclosed in a superconducting solenoidal magnet
providing a 1.0~T magnetic field. The solenoid is supported by an
octagonal flux-return yoke with resistive plate counter muon
identification modules interleaved with steel~\cite{detect}. 
The charged-particle momentum resolution at $1~{\rm GeV}/c$ is
$0.5\%$, and the 
${\rm d}E/{\rm d}x$
resolution is $6\%$ for electrons
from Bhabha scattering. The EMC measures photon energies with a
resolution of $2.5\%$ ($5\%$) at $1$~GeV in the barrel (end cap)
region. The time resolution in the TOF barrel region is 68~ps, while
that in the end cap region was 110~ps.  The end cap TOF
system was upgraded in 2015 using multigap resistive plate chamber
technology, providing a time resolution of 60~ps~\cite{etof}.
About 83\% of the data used here benefits from this upgrade.  
 
Simulated data samples produced with a {\sc geant4}-based~\cite{geant4} Monte Carlo~(MC) simulation 
software package, which includes
the BESIII detector response and geometric description,
are used to optimize the event selection criteria,
determine the detection efficiency, and estimate the backgrounds. 
For the signal process, we generate 100,000 $\ee\to\gamma\x$ events at each c.m.~energy, assuming an E1 radiative transition process which has been confirmed by BESIII data~\cite{gammax3872}.  
The $\x\to\pp\chico$ decay is described with the phase-space model in {\sc evtgen}~\cite{evtgen}. 
Initial-state-radiation (ISR) is simulated with {\sc kkmc}~\cite{kkmc}, by incorporating the $\sqrt{s}$-dependent production
cross section of $\ee\to\gamma\x$ into the program~\cite{BESIII-wjpsi}.
The maximum ISR photon energy is set according to  the production threshold of the $\gamma\x$ system.
Final-state-radiation is simulated with the {\sc photos} package~\cite{photos}.


The background contributions are investigated using an inclusive MC sample, which includes the production of open-charm processes, the ISR production of vector charmonium(-like) states, and the continuum processes incorporated in {\sc kkmc}.
All particle decays are modelled with the {\sc evtgen}~\cite{evtgen} using branching fractions taken from the Particle Data Group (PDG)~\cite{pdg} when available, and otherwise modelled with {\sc lundcharm}~\cite{lundcharm}. The equivalent luminosity of the inclusive MC sample is 40 times that of data at $\sqrt{s} = 4.178$ GeV, and is of equal size to data at other c.m.~energies.  A generic event-type analysis tool, TopoAna~\cite{topo}, is employed to study the backgrounds.


Charged tracks are required to be within a polar angle range of $|\rm{\cos\theta}|<0.93$,
where $\theta$ is defined with respect to the $z$-axis (the symmetry axis of the MDC). 
For each charged track, the distance of the closest approach to the interaction point (IP) must be less than 10\,cm along the $z$-axis, and less than 1\,cm
in the transverse plane.
The pions from the $\x$ decay and the leptons from the $\jpsi$ decay are kinematically well-separated.
Thus, charged tracks with momenta greater than 1.0~GeV/$c$ in the laboratory frame are taken as lepton candidates, while those with momenta less than 0.3~GeV/$c$ are taken as pion candidates due to limited phase space.
The energy deposition in the EMC of a lepton candidate is used to separate $e$ from $\mu$. 
Both $\mu$ candidates are required to have deposited energies less than 0.4~GeV, while both $e$ candidates are required to have deposited energies greater than 1.1~GeV.

Photon candidates are identified using showers in the EMC. 
The deposited energy of each shower must be more than 25~MeV in the barrel region ($|\cos \theta|< 0.80$) and more than 50~MeV in the end cap region ($0.86 <|\cos \theta|< 0.92$). 
To exclude showers that originate from charged tracks, the angle subtended by the EMC shower and the position of the closest charged track at the EMC must be greater than 10 degrees as measured from the IP. 
To suppress electronic noise and showers unrelated to the event, the difference between the EMC time of a shower 
and the event start time is required to be within 
[0, 700]\,ns.  
At least two good photon candidates are required in each event.


For an event with one soft pion undetected, the number of charged tracks ($\pi^\pm\ell^+\ell^-$) 
in the event is required to be three (referred to as a 3-track event).  
The four-momentum of the missing pion is obtained from four-momentum conservation and the initial beam kinematics.  
A one-constraint (1C) kinematic fit is then applied to each candidate event,
where the mass of the missing particle is constrained to the pion mass.
If there is more than one combination
 within an event due to additional photon candidates, 
we retain the two photons corresponding to the minimum $\chi^2$ from the 1C kinematic fit.
To further distinguish the radiative photon in $\EE\to\gamma_{\rm rad}\x$ and the photon from $\chico$ decay,
an extra constraint is added to constrain the $\gamma\LL$ invariant mass to the  $\chi_{c1}$ mass.
The combination with the minimum $\chi^2$ from the 2C kinematic fit is assigned as the correct combination,
and events with $\chi_{\rm 2C}^2<16$ are selected.

There are background events from the $\ee\to\gamma_{\rm ISR}\psip\to\gamma_{\rm ISR}\pp\jpsi$ process, together with one fake photon candidate.  
To remove such events, we require $\frac{E_{\gamma}^H-E_{\gamma}^L}{E_{\gamma}^H+E_{\gamma}^L}< 0.25$, where $E_{\gamma}^H$ and $E_{\gamma}^L$ are the energies of the higher and lower energy photons in an event, respectively. 
This criterion is very effective at rejecting background 
while retaining high signal efficiency.  
To reject $\gamma$ conversion ($\gamma\to\EE$) background events, where the converted electrons are misidentified as pions, particle identification~(PID) for charged pion, combining measurements of the specific ionization energy loss in the MDC and the flight time in the TOF to form likelihoods $\mathcal{L}(h)~(h=p,K,\pi)$ for each hadron $h$ hypothesis, is performed.
Tracks with momentum less than 0.3~GeV/$c$ must satisfy $\mathcal{L}(\pi)>\mathcal{L}(e)$. 
Background events from the $\ee\to\eta\jpsi\to\pp\piz\jpsi$ process are 
effectively vetoed by requiring $M(\gamma\gamma\pp) > 0.6$~GeV/$c^2$.


For events with both pions detected, the total number of charged tracks ($\pp\LL$) must be at least four (referred to as 4-track events), 
with exactly two oppositely-charged leptons, and at least two oppositely-charged pions. 
A four-constraint (4C) kinematic fit imposing four-momentum conservation is performed to each event. If there is more than one combination due to extra pion or extra photon candidates, 
we retain the one with the minimum $\chi^2$ value from the 4C kinematic fit.
An extra constraint is added to constrain $\gamma\LL$ to the $\chi_{c1}$ mass, to distinguish the photon from the $\chi_{c1}$ decay and the radiative photon associated with $\x$ production. 
The combination with the minimum $\chi^2$ from this 5C kinematic fit is assigned as the correct one.
Events with $\chi_{\rm 5C}^2<80$ are kept for further analysis.

There are $\psip$ background events coming from $\ee\to\pp\psip$ with $\psip\to \gamma\chi_{c1}/\pi^0\pi^0J/\psi$, 
and $\ee \to \gamma_{\rm ISR}\psip/\eta\psip$ with $\psip \to \pp \jpsi$. These background events are 
effectively vetoed by requiring the recoil mass of the $\pp$ system to satisfy 
$M^{\rm recoil}(\pip\pim)=\sqrt{(P_{\EE}-P_{\pp})^2}>3.704$~GeV/$c^2$, 
and the invariant mass $M(\pip\pim J/\psi) < 3.660$~GeV/$c^2$.
Here $P_{\EE}$ and $P_{\pp}$ are the four-momenta of the initial colliding beams and the $\pp$ pair, respectively.
The $\gamma$-conversion background events 
are effectively eliminated by requiring the opening angle of the pion pair to satisfy $\cos\theta_{\pip\pim} < 0.98$. 
Background events from the $\ee\to\eta\jpsi\to\pp\piz\jpsi$ process are rejected by requiring $M(\gamma\gamma\pp) > 0.59$~GeV/$c^2$.

\begin{figure}
	\centering
	\includegraphics[width=0.4\textwidth]{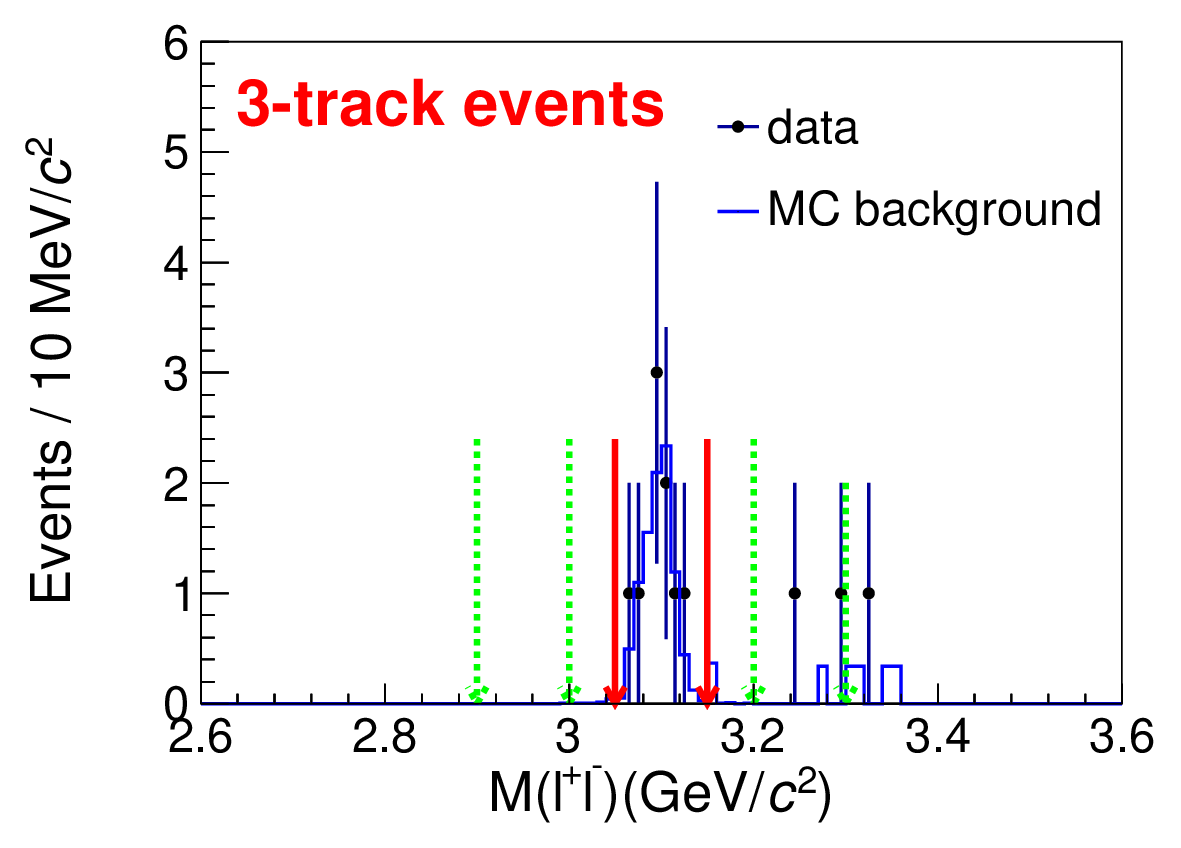}
	\includegraphics[width=0.4\textwidth]{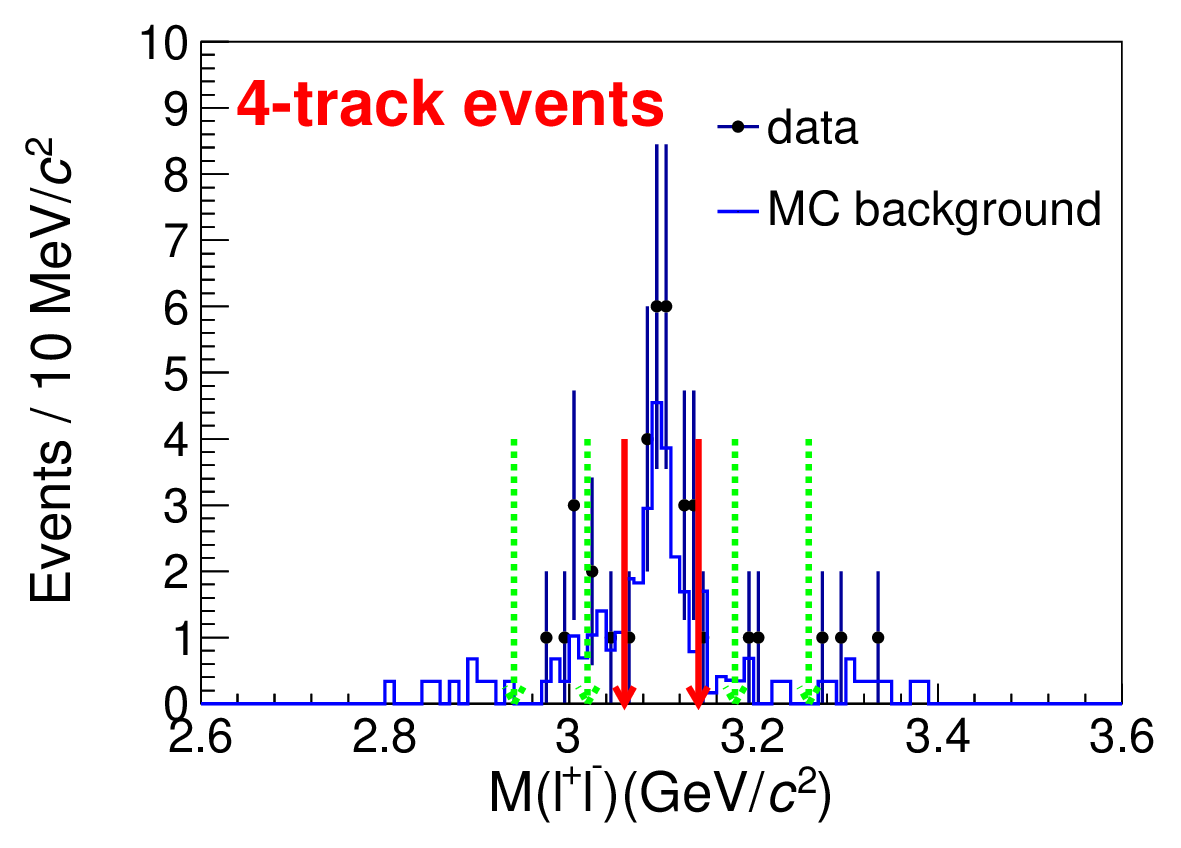}
	\vspace{-5mm}
	\caption{The distribution of $M(\LL)$ for data and MC background for 3-track events (top) and 4-track events (bottom). The dots with error bars are the full data. The blue histogram represents the MC background, which come from the contributions of $\eta'\jpsi$, $\pp\psi(3823)$, $\gamma_{\rm ISR}\psi'$ and continue background. The area between the red solid arrows is the signal area of $\jpsi$, and the area between the green dotted arrows is the sideband area of $\jpsi$.}
	\label{mll}
\end{figure}

The $M(\LL)$ mass distribution after performing the above selections is shown in Fig.~\ref{mll}. To select signal candidates that contain a $\jpsi$ resonance, we define
[3.05, 3.15]~GeV/$c^2$ (for 3-track events), and [3.06, 3.14]~GeV/$c^2$ (for 4-track events) 
as the $\jpsi$ mass windows. 
The non-$\jpsi$ background contribution is estimated by the events in the $\jpsi$ mass sideband regions, which are defined as [2.90, 3.00]~GeV/$c^2$ or [3.20, 3.30]~GeV/$c^2$ (for 3-track events), and [2.94, 3.02]~GeV/$c^2$ or [3.18, 3.26]~GeV/$c^2$ (for 4-track events), respectively. 

\begin{figure}
	\centering
	\includegraphics[width=0.4\textwidth]{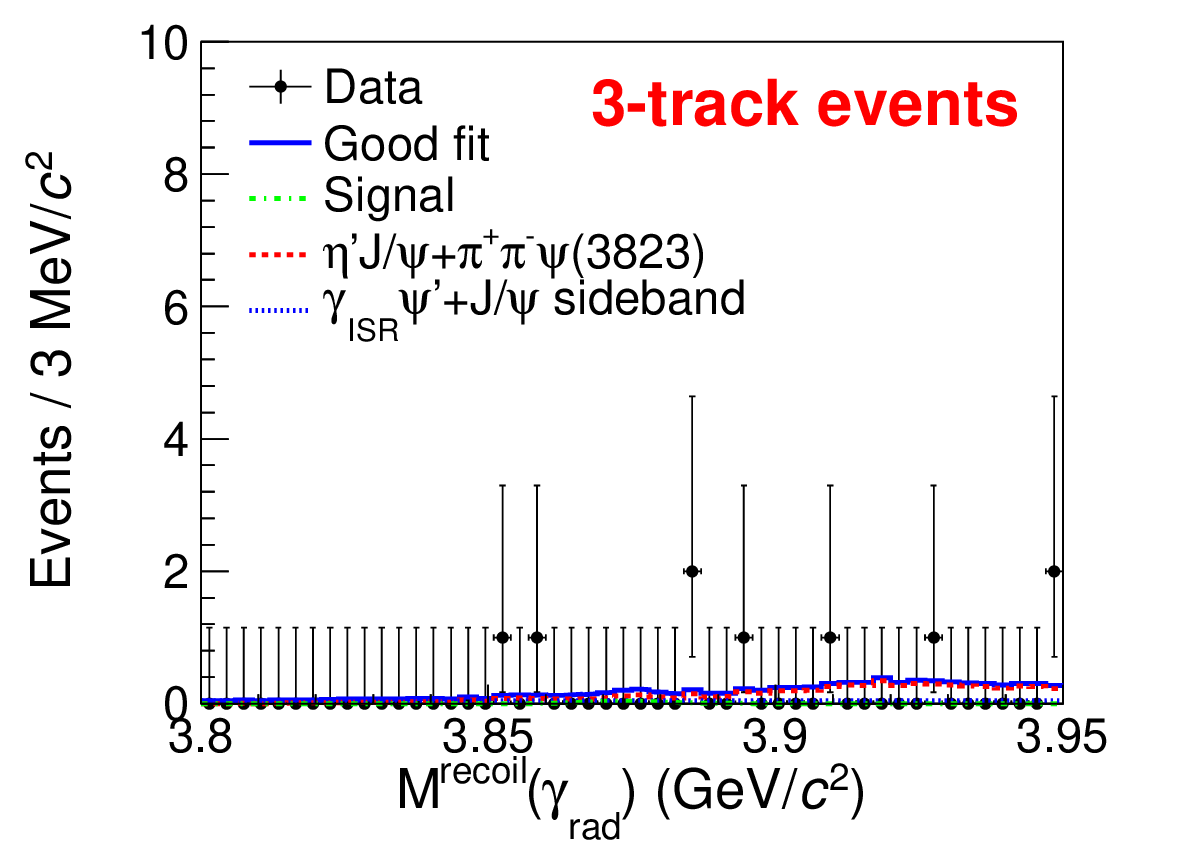}\\
\includegraphics[width=0.4\textwidth]{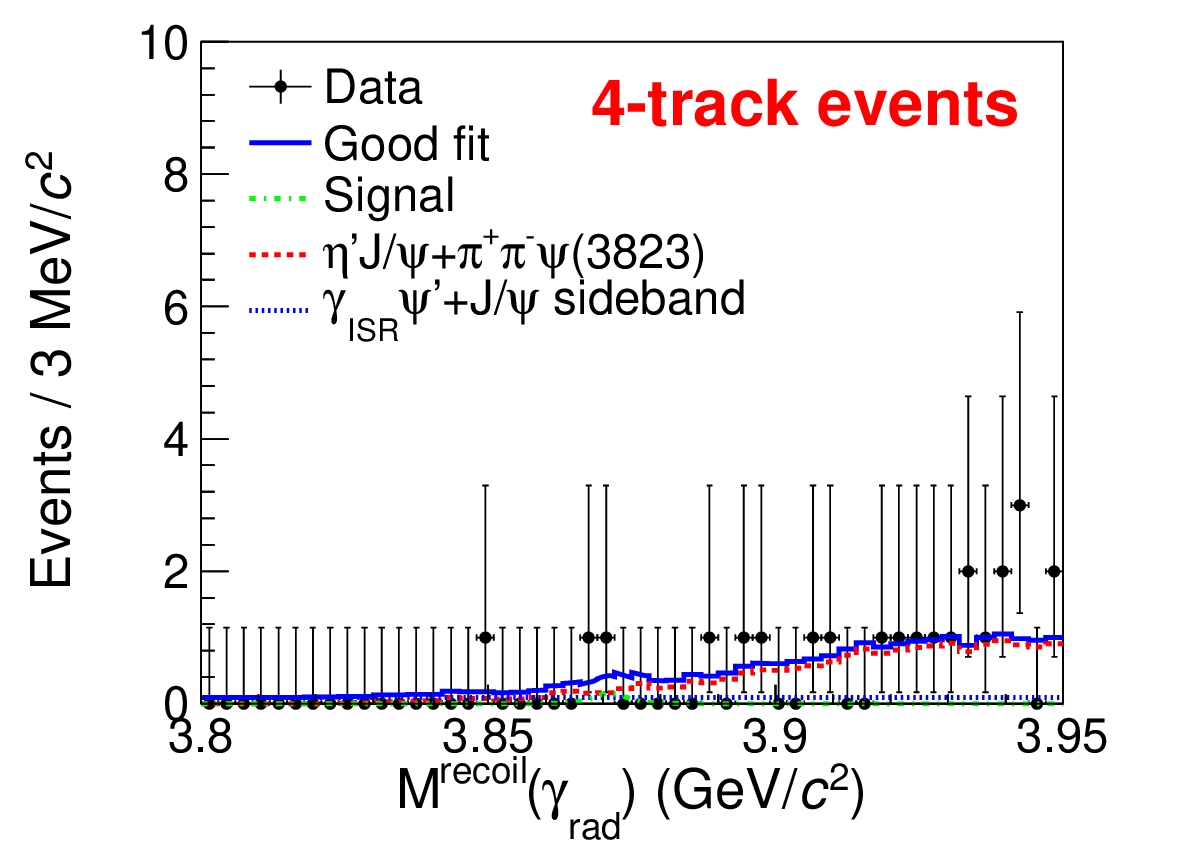}	
\vspace{-5mm}
	\caption{Result of the simultaneous fit to the $M^{\rm recoil}(\gamma_{\rm rad})$ distributions for 
	3-track events (top) and  4-track events (bottom).
	The dots with error bars are the full data, the blue solid curves represent the total fit, 
	the green dash-dotted curves are the signal contribution, and 
	the blue dotted curves and red dashed curves are the background contributions.}
	\label{simfit}
\end{figure}

Figure~\ref{simfit} shows the recoil mass distribution 
of the radiative photon $M^{\rm recoil}(\gamma_{\rm rad})=\sqrt{(P_{\EE}-P_{\gamma_{\rm rad}})^2}$ from the full data set after imposing the above requirements. Here $P_{\gamma_{\rm rad}}$ is the four-momenta of the radiative photon $\gamma_{\rm rad}$.
The background level is very low, and no obvious $\x$ signal is observed.
Possible remaining background contributions mainly come from the 
$\ee \to \eta^\prime J/\psi \to \pip\pim\eta\jpsi$ and $\ee \to \pip\pim \psi_2(3823) \to \gamma \pp\chi_{c1}$ processes. 
These background contributions have been well-studied by BESIII~\cite{etapjpsi, ppx3823}
and can be reliably simulated, as shown in Fig.~\ref{simfit}.
According to a study of the $\jpsi$ mass sideband events and inclusive MC events, the non-$\jpsi$ background  and $\gamma_{\rm ISR}\psip$ contributions are found to be small
and only produce a flat distribution in the $\x$ signal region.

To extract the signal yield, a simultaneous unbinned maximum likelihood fit is performed to the $M^{\rm recoil}(\gamma_{\rm rad})$ 
distributions for the 3-track and 4-track events, as shown in Fig.~\ref{simfit}.
In the simultaneous fit, the ratio between the 3-track and 4-track events signal yields is fixed according to the detection efficiencies.
The signal probability density function in the fit is represented by the MC-simulated $\x$ shape convolved with a Gaussian function,
which accounts for the difference in mass resolution between data and MC simulation.
The parameters of the $\x$ resonance in the simulation are taken from PDG~\cite{pdg}, and those of
the convolved Gaussian are fixed according to a study of the
$\ee\to\gamma_{\rm ISR}\psip\to\gamma_{\rm ISR}\ppjpsi$ control sample.
The background in the fit consists of two components.
One is the simulated contribution from the $\ee \to \eta' J/\psi$ and $\ee \to \pip\pim \psi_2(3823)$ processes, 
which is normalized according to their measured cross sections~\cite{etapjpsi, ppx3823}. 
The other is a mass-independent term, which represents the contributions from the simulated $\gamma_{\rm ISR}\psip$
background events and continuum background events estimated from $\jpsi$ mass sideband data. 

Since no obvious $\x\to\pp\chico$ signal is observed in the $M^{\rm recoil}(\gamma_{\rm rad})$ distributions, 
we estimate the upper limit (U.L.) for the produced number of $\x\to\pp\chico$ yield $N^{\rm sig}(\pp\chico)$ 
after efficiency correction.
Based on the Bayesian method~\cite{Bayes}, the U.L. at a 90\% confidence level (C.L.) 
is set to be $N^{\rm sig}(\pp\chico)<17.8$
by scanning the likelihood curve in the simultaneous fit with all the additive systematic uncertainties taken into account
.
To take into account the multiplicative systematic uncertainty, the likelihood curve is further convolved by a Gaussian with a width parameter equal to the total multiplicative systematic uncertainty, 13.2$\%$ ($\int_{0}^{+\infty} \mathcal{L}(N^{\prime}_{sig}) e^{-\frac{(N_{sig} - N^{\prime}_{sig})^2}{2(0.132 N_{sig})^2}} dN^{\prime}_{sig}$, where $N^{(\prime)}_{sig}$ is the signal yield, $\mathcal{L}(N^{\prime}_{sig})$ is the likelihood curve).
	And the systematic uncertainties are discussed below.
The most conservative estimate of U.L. after considering all systematic uncertainty is 
$N^{\rm sig}(\pp\chico)<18.5$ at a 90$\%$ C.L.



The relative branching ratio $\mathcal{R}\equiv\frac{\mathcal{B}[\x \to \pip\pim\chi_{c1}]}{\mathcal{B}[\x \to \pip\pim J/\psi]}$ is calculated as
\begin{linenomath*}  
\begin{equation}
	\begin{aligned}
		\mathcal{R} 
		&= \frac{N^{\rm sig}(\pip\pim\chi_{c1})\epsilon^{\rm ave}(\pip\pim J/\psi)}{N^{\rm obs}(\pip\pim J/\psi)\mathcal{B}(\chi_{c1} \to \gamma J/\psi)},
	\end{aligned}
\end{equation}
\end{linenomath*}
where $N^{\rm obs}(\pip\pim J/\psi)=86.3^{+10.5}_{-9.8}$ is the signal yield of $\x\to\pp\jpsi$
from the same data set,
$\epsilon^{\rm ave}(\pip\pim J/\psi)=0.287$ is the weighted average efficiency of $\x\to\pp\jpsi$ events~\cite{gammax3872}, 
and $\mathcal{B}(\chi_{c1} \to \gamma J/\psi)=0.343$ is the branching fraction of $\chico\to\gamma\jpsi$~\cite{pdg}.
The U.L. for the ratio is determined to be $\mathcal{R}<0.18$ at a 90$\%$ C.L.



The U.L. for the product of the cross section $\sigma[\ee\to\gamma\x]$ 
and the branching fraction $\mathcal{B}[\x\to\pp\chico]$ is calculated as 
\begin{linenomath*}  
\begin{eqnarray}\notag
\sigma[\ee\to\gamma\x] \, \mathcal{B}[\x\to\pp\chico] \\ = \frac{N^{\rm U.L.}}{\mathcal{L}_{\rm int}(1+\delta)\epsilon_{\rm tot}\mathcal{B}},
\end{eqnarray} 
\end{linenomath*}
where $\mathcal{L}_{\rm int}$ is the integrated luminosity, $(1+\delta)$ is the radiative correction factor calculated by the {\sc kkmc} program~\cite{kkmc}.
 The symbol $\epsilon_{\rm tot}$ is the sum of selection efficiencies for 3-track events and 4-track events, and $\mathcal{B}\equiv\mathcal{B}(\chico\to\gamma\jpsi) \, \mathcal{B}(\jpsi\to\LL)$ is a product of the corresponding branching fractions.
$N^{\rm U.L.}$ is the U.L. at a $90\%$ C.L. for the signal yield at each c.m.~energy, which is determined by counting the number of events in the $\x$ signal region [3.86, 3.88]~GeV/$c^2$, due to the limited statistics. The background has been subtracted, which is estimated by using the $\x$ sidebands [3.81, 3.84]~GeV/$c^2$ and [3.91, 3.94]~GeV/$c^2$. The distributions of $M^{\rm recoil}(\gamma_{\rm rad})$ at each c.m.~energy are shown in Appendix~\ref{rmgy-all} Fig.~\ref{rmgy}. The $N^{\rm U.L.}$ at each c.m.~energy is calculated using a frequentist method with an unbounded profile likelihood treatment by assuming the numbers of observed events in the $\x$ signal and sideband regions follow a Poisson distribution~\cite{TRolke}.
Table~\ref{count_UL} summarizes the results related to the $\sigma[\ee\to\gamma\x] \, \mathcal{B}[\x\to\pp\chico]$ measurement.

\begin{table*}[ht]
	\caption{ 
 Summary of the integrated luminosities ($\mathcal{L}_{\rm int}$) of data, the total efficiency of 3-track events and 4-track events ($\epsilon_{\rm tot}$), the ISR correction factor (1+$\delta$), the number of observed events, $N_{\rm obs}$, the expected number of background events, $N_{\rm bkg}$, in the $\x$ signal region for each sample, the obtained 90$\%$ C.L. upper limit for the $\x$ signal yields, $N^{\rm U.L.}$, and the product $\sigma[\ee\to\gamma\x] \, \mathcal{B}[\x\to\pp\chico]$ at each c.m.~energy, denoted as $(\sigma\mathcal{B})^{\rm U.L.}$, in pb. The multiplicative systematic uncertainties, denoted $\Delta$ in $\%$, have been taken into account.
}
	\centering
	\begin{tabular}{c c c c c c c c c} 
		\hline 
		\hline
		$\sqrt{s}$~(GeV) & $\mathcal{L}_{\rm int}$~(pb$^{-1}$) & $\epsilon_{\rm tot}~(\%)$ &  $(1+\delta)$ & $N_{\rm obs}$ & $N_{\rm bkg}$ & $N^{\rm U.L.}$ & $(\sigma\mathcal{B})^{\rm U.L.}$ & $\Delta~(\%)$\\
		\hline
4.158 & 408.2 & 32.6 & 0.78 & 0 & 0.3 & 1.8 & 0.50 & 5.3 \\
4.178 & 3194.5 & 32.3 & 0.78 & 2 & 0.3 & 5.0 & 0.18 & 5.4 \\
4.189 & 526.7 & 32.7 & 0.78 & 0 & 0.2 & 1.8 & 0.39 & 5.5 \\
4.199 & 526.0 & 32.9 & 0.79 & 0 & 0.2 & 1.8 & 0.38 & 5.5 \\
4.209 & 517.1 & 32.3 & 0.81 & 0 & 0.0 & 2.0 & 0.43 & 5.5 \\
4.219 & 514.6 & 32.3 & 0.84 & 0 & 0.2 & 1.8 & 0.38 & 5.5 \\
4.226 & 1056.4 & 32.8 & 0.86 & 0 & 0.3 & 1.7 & 0.17 & 5.4 \\
4.236 & 530.3 & 32.0 & 0.89 & 0 & 0.1 & 1.9 & 0.37 & 5.4 \\
4.244 & 538.1 & 31.6 & 0.92 & 0 & 0.1 & 1.9 & 0.37 & 5.4 \\
4.258 & 828.4 & 30.9 & 0.97 & 0 & 0.0 & 2.0 & 0.25 & 5.4 \\
4.267 & 531.1 & 30.0 & 1.00 & 0 & 0.0 & 2.0 & 0.40 & 5.4 \\
4.278 & 175.7 & 28.8 & 1.05 & 0 & 0.0 & 2.0 & 1.24 & 5.6 \\
4.288 & 502.4 & 28.3 & 1.09 & 0 & 0.1 & 1.9 & 0.43 & 5.5 \\
4.312 & 501.1 & 26.2 & 1.19 & 0 & 0.0 & 2.0 & 0.44 & 5.5 \\
4.338 & 504.9 & 23.9 & 1.30 & 0 & 0.0 & 2.0 & 0.43 & 5.5 \\
		\hline 
		\hline
	\end{tabular}
	\label{count_UL}  
\end{table*}


For the $\mathcal{R}$ measurement, many common systematic uncertainty sources
cancel, including those arising from the luminosity, the lepton reconstruction
efficiency, the kinematic fit, the branching fraction of $\mathcal{B}(\jpsi\to\LL)$, and the ISR correction factor.
 The non-canceling systematic uncertainties consists of multiplicative and additive systematic uncertainties. The multiplicative are derived from the efficiencies for photon detection, pion tracking, PID, the branching fraction of $\chico\to\gamma\jpsi$,  the MC decay model, and those from $\x\to\pp\jpsi$ channel. 
The uncertainty for photon detection is 1.0$\%$ per photon~\cite{uncertaintyofphoton}.
Since the efficiency for one photon does not cancel in  $\mathcal{R}$,
1.0\% is taken as the systematic uncertainty for photon detection.  
The pion pair in the search channel of $\x\to\pp\chico$ has a relatively 
low momentum compared to that of $\x\to\ppjpsi$.
Thus the uncertainty of pion tracking does not cancel, 
and the uncertainty is estimated 
to be 1.0\% per pion~\cite{syserrorofpi}.  
In the search channel, PID is applied for the pion candidate in 3-track event
selection.  The systematic uncertainty is assigned as 1.0$\%$ per pion.
The systematic uncertainty related to the branching fraction 
$\mathcal{B}(\chico\to\gamma\jpsi)$ is taken from PDG~\cite{pdg}. 
In the nominal analysis, the $\x\to\pp\chico$ signal MC events are generated using a phase-space model.
Assuming the $\pip\pim$ system is dominated by a $\sigma(500)$ resonance, we generate alternative signal MC events for both $S$-wave and $D$-wave $\x$ decays, and the maximum difference to the nominal efficiency is taken as systematic uncertainty from an MC decay model. 
%
%
The systematic uncertainty from the $\x \to \pp\jpsi$ channel is quoted from Ref.~\cite{gammax3872}, 
including sources related to the pion tracking, signal parametrization, background shape, and the statistical uncertainty of the signal yield $N^{\rm sig}(\pp\jpsi)$.

Assuming that all of the multiplicative systematics are independent, 
the total multiplicative systematic uncertainty for the $\mathcal{R}$ ratio measurement is obtained by adding all these individual uncertainties in quadrature,
resulting in 13.2\%, as shown in Table~\ref{systematical error for ratio}.

\begin{table}
	\caption{Summary of the multiplicative systematic uncertainties (in $\%$) for the $\mathcal{R}$ measurement.}
	\begin{tabular}{cccc} 
		\hline
		\hline
		Source & 3-track ~&~ 4-track ~&~  Combined \\
		\hline
		Photon detection efficiency & 1.0 ~&~1.0~&~ 1.0\\
		Pion tracking efficiency & 1.0 ~&~ 2.0 ~&~ 1.6 \\
		PID & 1.0 ~&~ ... ~&~ 0.5 \\
		$\mathcal{B}(\chico\to\gamma\jpsi)$ & 3.0 ~&~3.0~&~ 3.0\\
		MC decay model & 2.8 ~&~2.8~&~ 2.8\\
		$\x\to\pip\pim\jpsi$ & ~&~ ~&~~12.4\\
		\hline
		Total & ~&~ ~&~~13.2\\
		\hline 
		\hline
	\end{tabular}
	\label{systematical error for ratio}  
\end{table}	

Other additive uncertainties including backgrounds, the fit range and the signal shape  affect $N^{\rm sig}(\pp\chico)$ directly.  
The uncertainty due to backgrounds in the $\x\to\pp\chico$ channel
is investigated by floating the background contributions from $\eta^\prime J/\psi$ and $\pip\pim \psi_2(3823)$ which were previously fixed based on MC studies.  
The uncertainty associated with the fit range is determined
by varying the fit range within $\pm 10$~MeV. The uncertainty due to
the signal shape is considered by varying the resolution of the convolved
Gaussian within $\pm 1 \sigma$. The most conservative $N^{\rm sig}(\pp\chico)$
	from the combined effects of these additive sources is taken as the final result.
%


The systematic uncertainties in the $\sigma[\ee\to\gamma\x] \, \mathcal{B}[\x\to\pp\chico]$ measurement include those from the luminosity,  photon detection, tracking efficiency,  PID, $\jpsi$ mass window, kinematic fit, MC decay model, 
radiative correction and branching fraction.
The luminosity is measured using large angle Bhabha events, with an uncertainty of $0.66\%$~\cite{lum}.
The systematic uncertainties related to the photon detection, PID, MC decay model and branching fraction of $\chico\to\gamma\jpsi$ are the same as those in the $\mathcal{R}$ measurement.

The uncertainty of tracking efficiency for the high-momentum leptons is 1.0\% per track~\cite{erroroflep}. By requiring at least one pion to
	be detected, the pion detection efficiency is very high and the
	uncertainty is negligible. The uncertainty of $\mathcal{B}(\jpsi\to\LL)$ is 0.6\%, quoted from PDG~\cite{pdg}.
The uncertainties caused by the $\jpsi$ mass window are studied with a control sample of $\ee\to\eta \jpsi$ events, 
with $\eta\to\pp\piz$, resulting in $0.55\%$ and $0.04\%$ for the 3-track events and 4-track events, respectively.
A track helix parameter correction method is applied to the signal MC events~\cite{kf-correction} in the kinematic fit. 
The difference in efficiencies with and without the correction is assigned as the systematic uncertainty from the kinematic fit.
To estimate the systematic uncertainty from radiative corrections (i.e., due to the uncertainty of the $\ee\to\gamma\x$ cross section line shape~\cite{BESIII-wjpsi}), we sample 300 line shapes using the resonance parameters ($M=4200.6^{+7.9}_{-13.3}\pm3.0$ MeV/$c^{2}$  and $\Gamma=115^{+38}_{-26}\pm12$ MeV) within uncertainties, including their correlation, taken from Ref. \cite{BESIII-wjpsi}.  
A weight method~\cite{sunradcor} is used to get the distribution of $(1+\delta)\epsilon$, 
and the standard deviation of $(1+\delta)\epsilon$ is estimated as the systematic uncertainty from radiative correction.

Assuming that all the sources are independent, the total systematic uncertainty for $\sigma[\ee\to\gamma\x] \, \mathcal{B}[\x\to\pp\chico]$ measurement at each c.m.~energy is calculated by adding them in quadrature, as listed in the last column of Table~\ref{count_UL}.
Table~\ref{systematical error for xs} in Appendix~\ref{sys-xs} summarizes the systematic uncertainties 
for the $\sigma[\ee\to\gamma\x] \, \mathcal{B}[\x\to\pp\chico]$ measurement.


In summary, with a data sample corresponding to an integrated luminosity of 10.9~fb$^{-1}$ collected with the BESIII detector at c.m.~energies ranging from 4.16 to 4.34~GeV, the process $\ee\to\gamma\x$ with $\x\to\pp\chico$ is studied, and no obvious signal is found. 
A 90\% C.L. upper limit on the $\sigma[\EE\to\gamma\x] \, \mathcal{B}[\x\to\pp\chico]$ is set at each c.m.~energy.  
We also set a limit on the ratio of branching fractions 
of $\mathcal{R}=\frac{\mathcal{B}[\x \to \pip\pim\chi_{c1}]}{\mathcal{B}[\x \to \pip\pim J/\psi]}<0.18$ at a 90$\%$ confidence level, 
which is consistent with the measurement from the Belle Collaboration~\cite{Belle:chico,Belle:ppjpsi}. 
Considering $\frac{\mathcal{B}[\x \to \piz\chi_{c1}]}{\mathcal{B}[\x \to \pip\pim J/\psi]}=
0.88^{+0.33}_{-0.27}\pm 0.10$~\cite{BESIII-pi0chic1}, the relative decay width 
$\frac{\Gamma[\x \to \chi_{c1}\piz]}{\Gamma[\x \to \chi_{c1}\pip\pim]}>5$ is determined, which is two orders of magnitude greater than that 
expected under a pure charmonium $2^3P_1$ assumption for the $\x$~\cite{pi0/pipi}.
These measurements favor the non-conventional charmonium nature of the $\x$ state~\cite{X3872-mole}, 
and help to constrain the $c\bar{c}$ core component in the $\x$ wave function~\cite{mixture}.


\textbf{Acknowledgement}

The BESIII Collaboration thanks the staff of BEPCII and the IHEP computing center for their strong support. This work is supported in part by National Key R\&D Program of China under Contracts Nos. 2020YFA0406300, 2020YFA0406400; National Natural Science Foundation of China (NSFC) under Contracts Nos. 11975141, 11635010, 11735014, 11835012, 11935015, 11935016, 11935018, 11961141012, 12025502, 12035009, 12035013, 12061131003, 12192260, 12192261, 12192262, 12192263, 12192264, 12192265, 12221005, 12225509, 12235017; the Chinese Academy of Sciences (CAS) Large-Scale Scientific Facility Program; the CAS Center for Excellence in Particle Physics (CCEPP); Joint Large-Scale Scientific Facility Funds of the NSFC and CAS under Contract No. U1832207; CAS Key Research Program of Frontier Sciences under Contracts Nos. QYZDJ-SSW-SLH003, QYZDJ-SSW-SLH040; 100 Talents Program of CAS; Project ZR2022JQ02 supported by Shandong Provincial Natural Science Foundation; Supported by the China Postdoctoral Science Foundation under Grant No. 2023M742100; The Institute of Nuclear and Particle Physics (INPAC) and Shanghai Key Laboratory for Particle Physics and Cosmology; European Union's Horizon 2020 research and innovation programme under Marie Sklodowska-Curie grant agreement under Contract No. 894790; German Research Foundation DFG under Contracts Nos. 455635585, Collaborative Research Center CRC 1044, FOR5327, GRK 2149; Istituto Nazionale di Fisica Nucleare, Italy; Ministry of Development of Turkey under Contract No. DPT2006K-120470; National Research Foundation of Korea under Contract No. NRF-2022R1A2C1092335; National Science and Technology fund of Mongolia; National Science Research and Innovation Fund (NSRF) via the Program Management Unit for Human Resources \& Institutional Development, Research and Innovation of Thailand under Contract No. B16F640076; Polish National Science Centre under Contract No. 2019/35/O/ST2/02907; The Swedish Research Council; U. S. Department of Energy under Contract No. DE-FG02-05ER41374


%
%

\clearpage
\onecolumngrid
\appendix

\section{The distribution of $M^{\rm recoil}(\gamma_{\rm rad})$ at each c.m.~energy}
\label{rmgy-all}
\begin{figure}[H]
	\centering
	\includegraphics[width=0.3\textwidth]{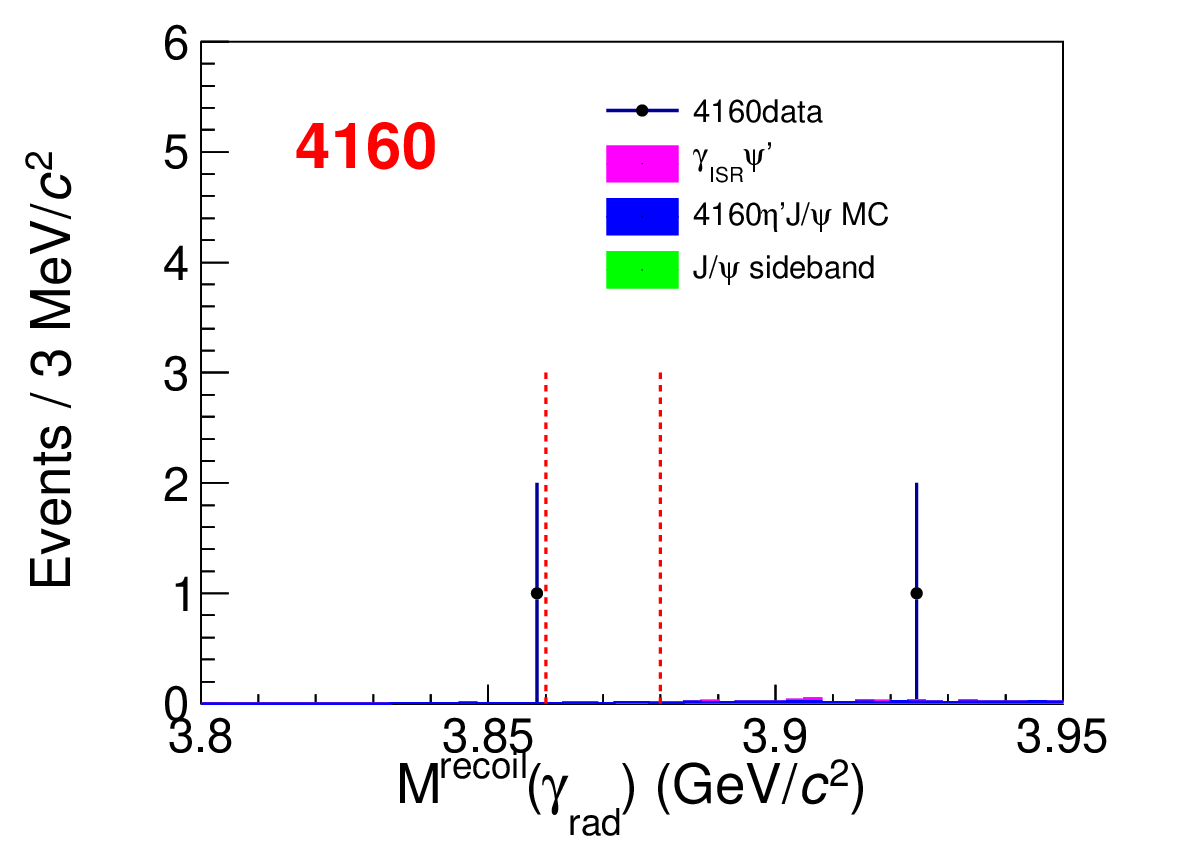}\hspace{5pt}
    \includegraphics[width=0.3\textwidth]{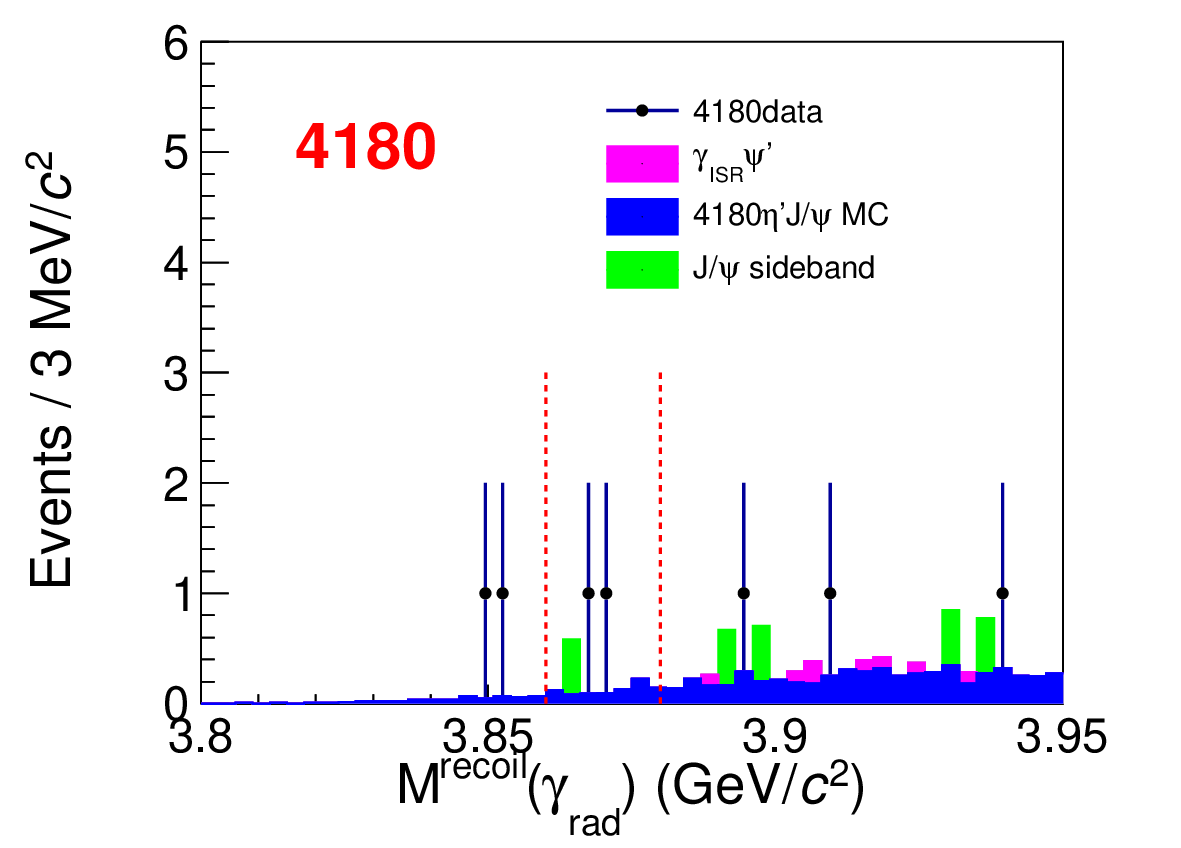}\hspace{5pt}
    \includegraphics[width=0.3\textwidth]{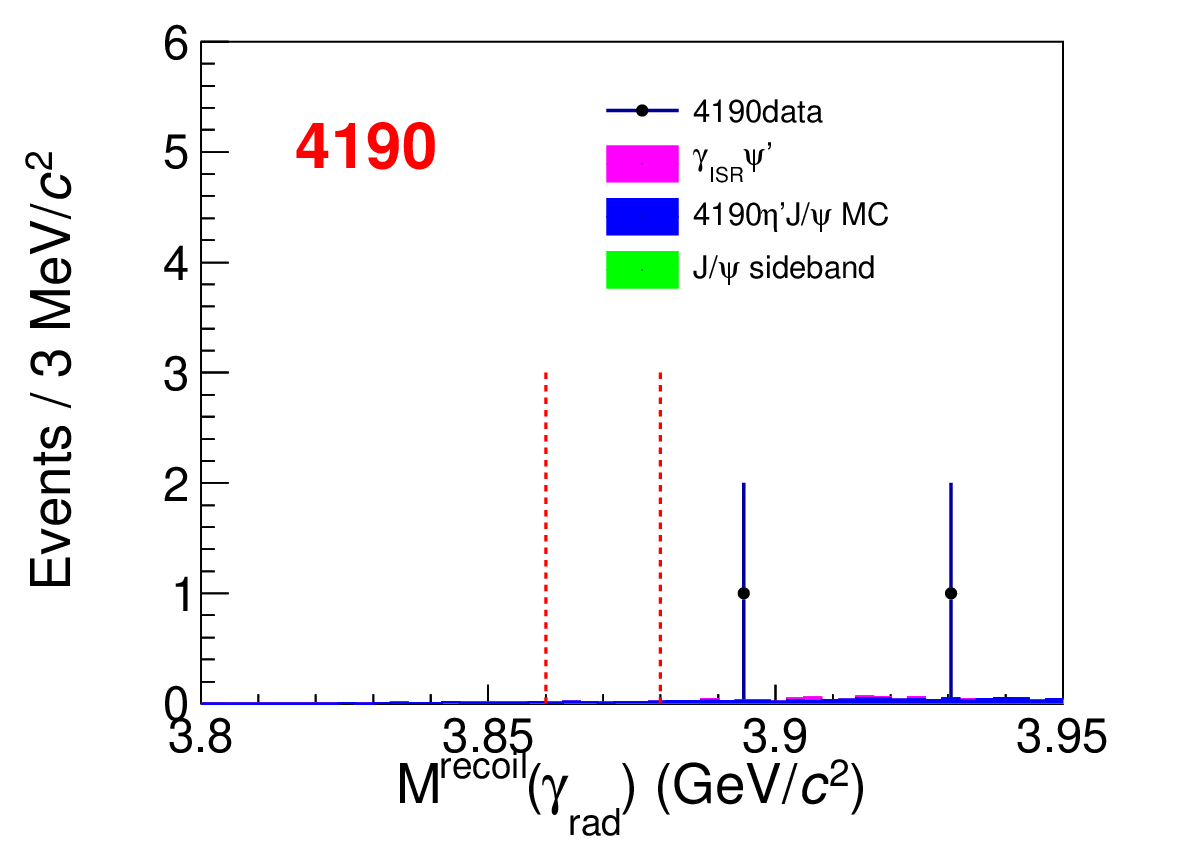}\hspace{5pt}
    \includegraphics[width=0.3\textwidth]{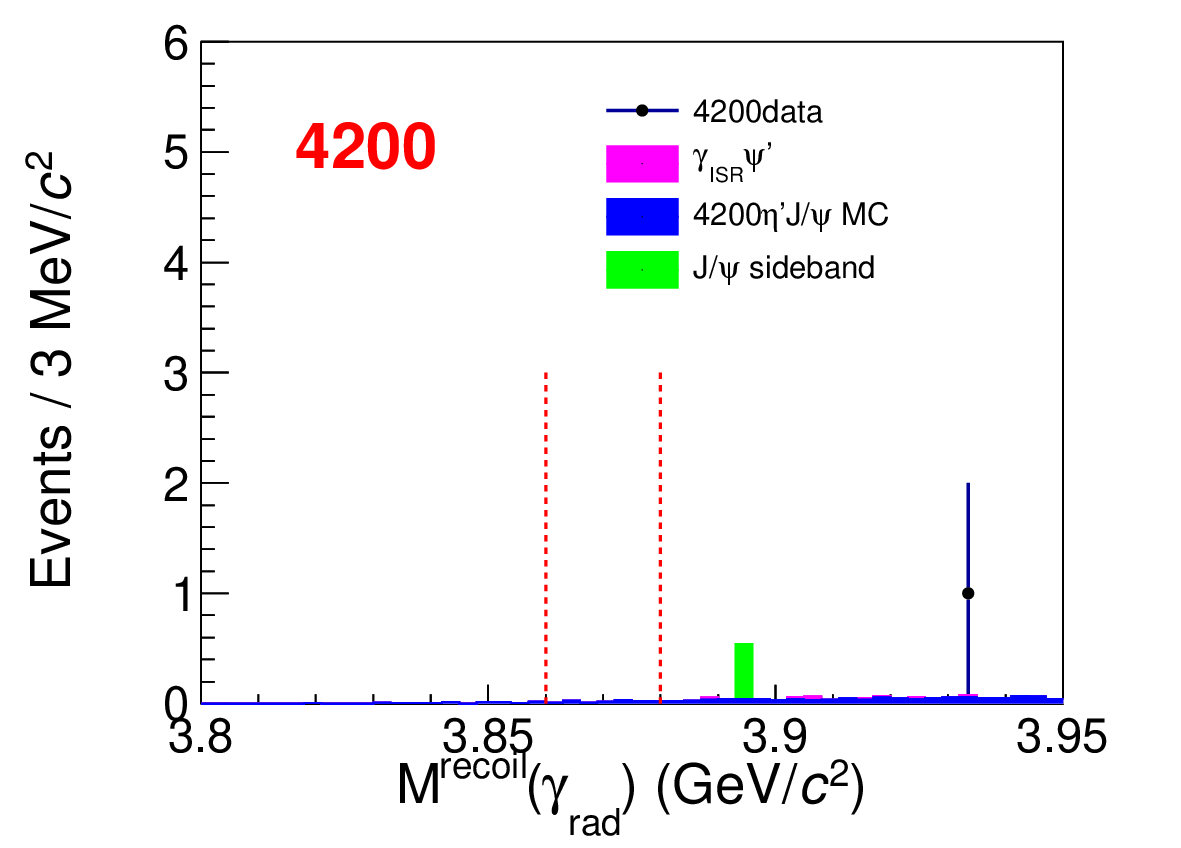}\hspace{5pt}
    \includegraphics[width=0.3\textwidth]{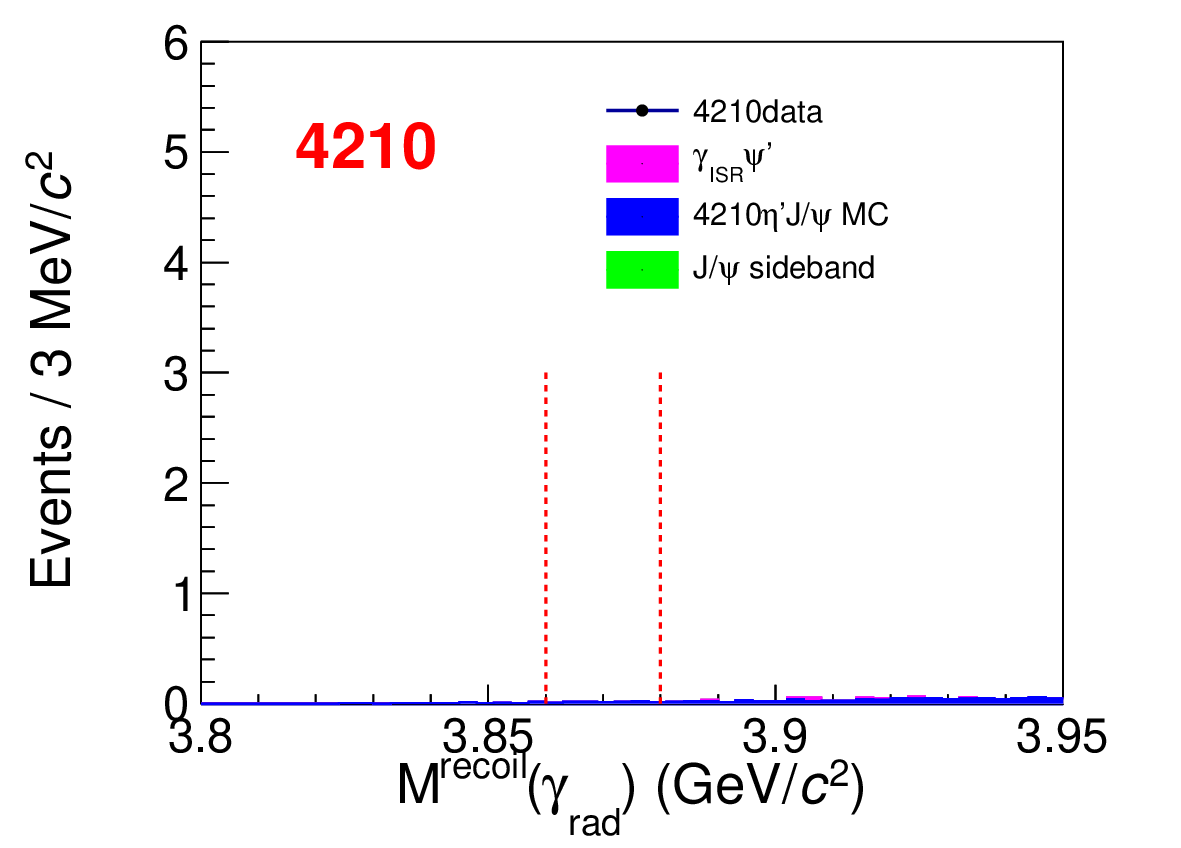}\hspace{5pt}
    \includegraphics[width=0.3\textwidth]{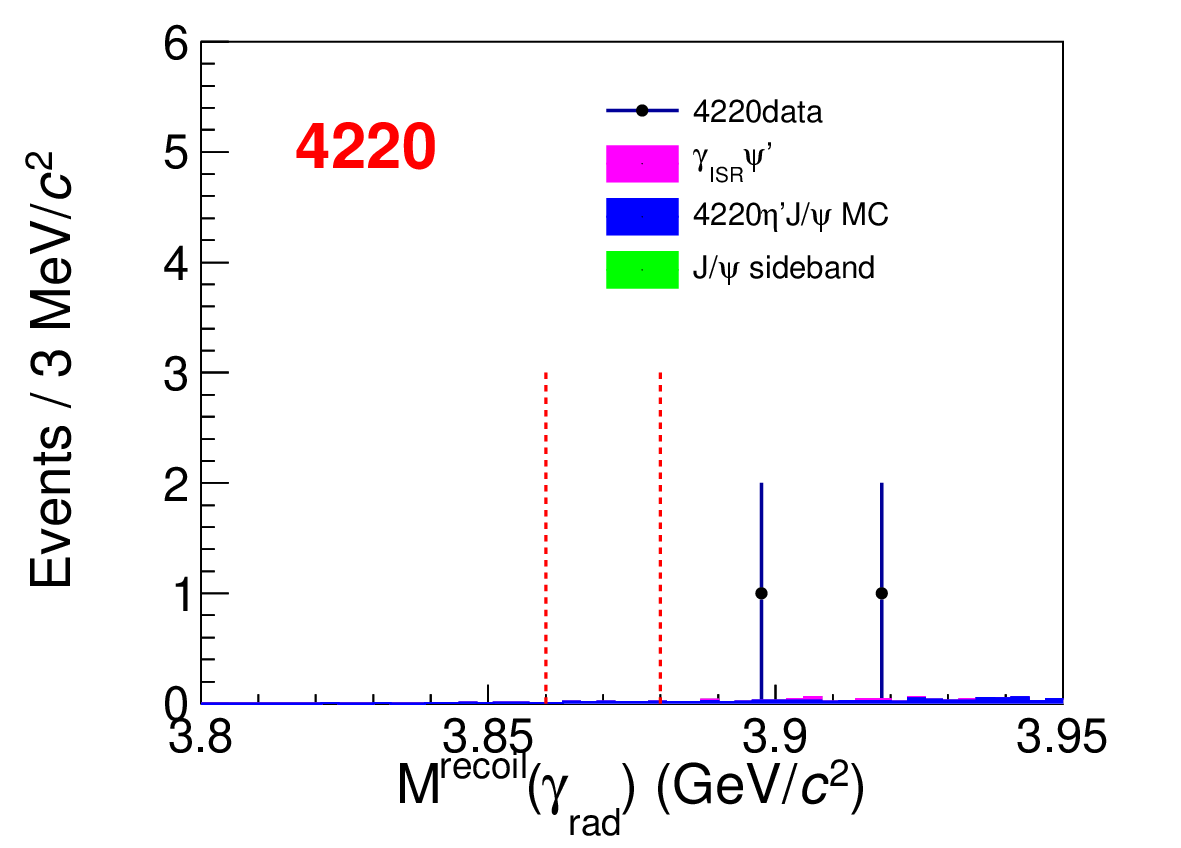}\hspace{5pt}
    \includegraphics[width=0.3\textwidth]{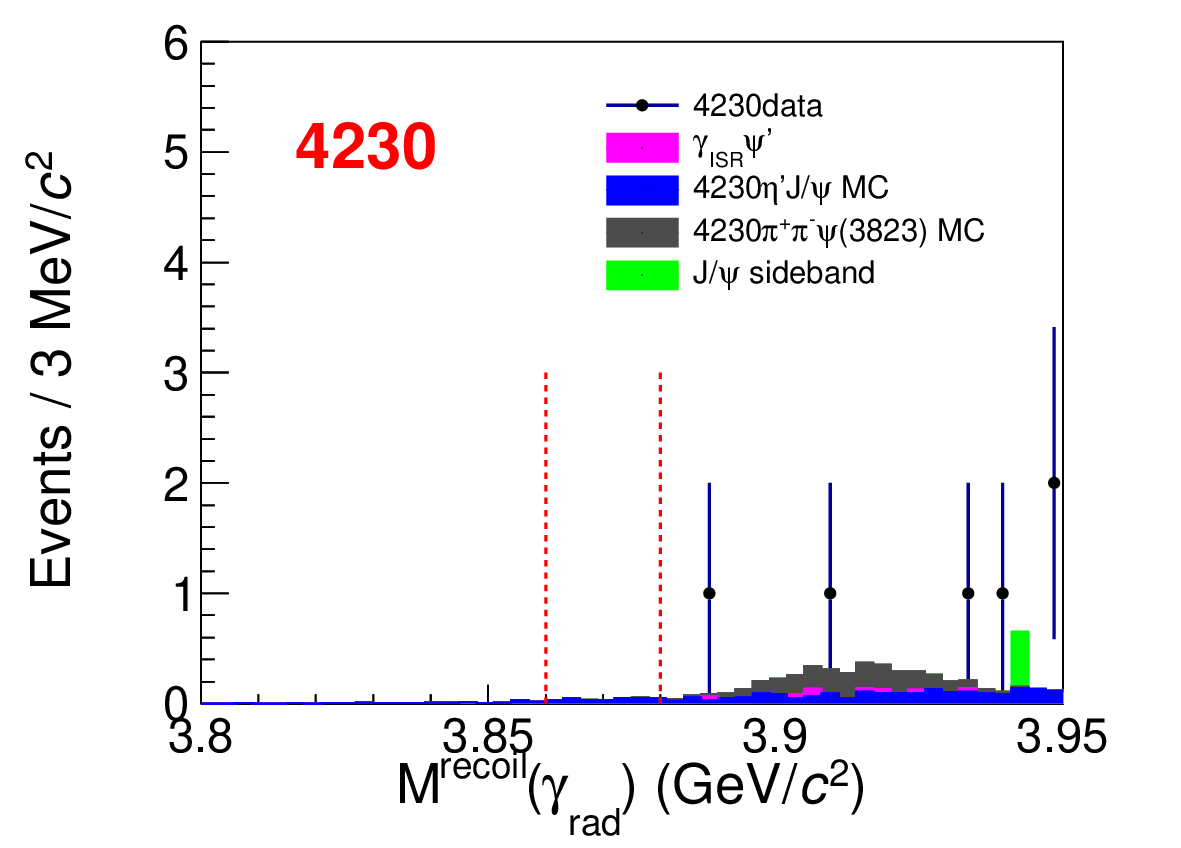}\hspace{5pt}
    \includegraphics[width=0.3\textwidth]{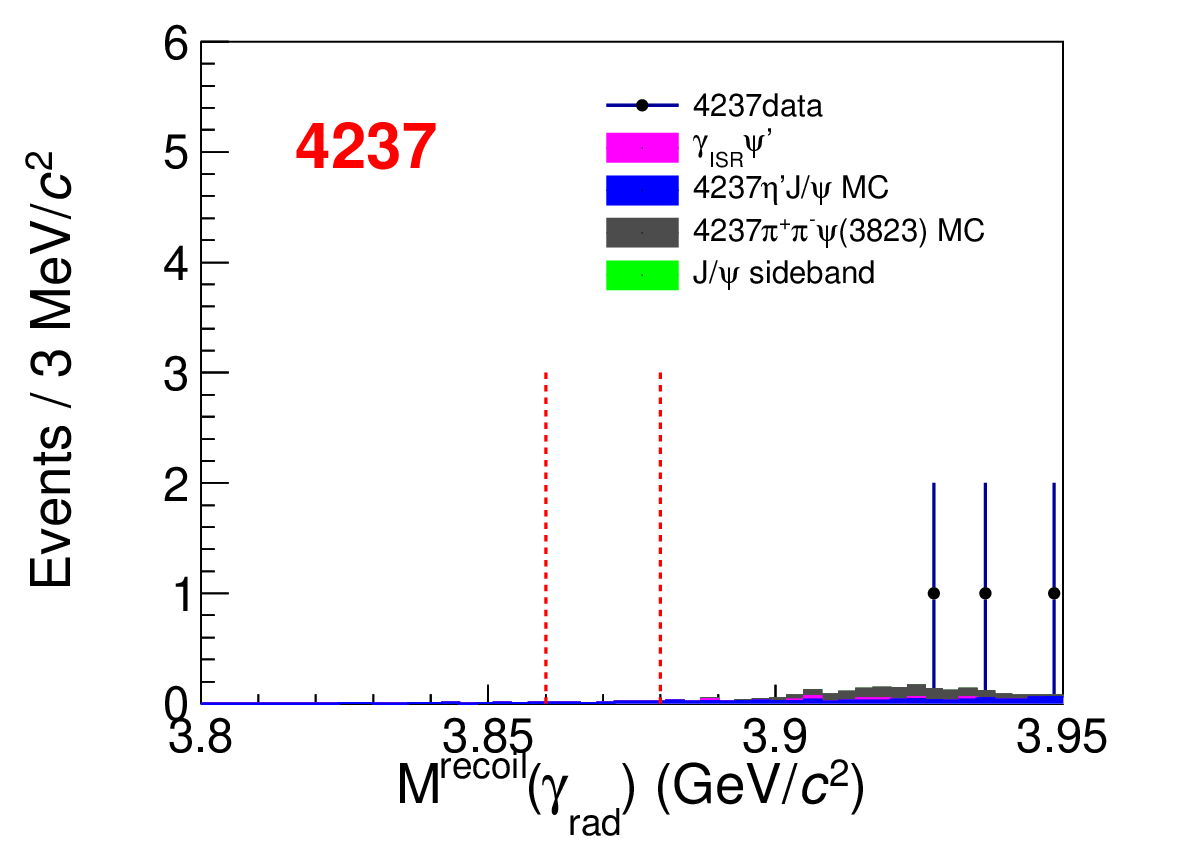}\hspace{5pt}
    \includegraphics[width=0.3\textwidth]{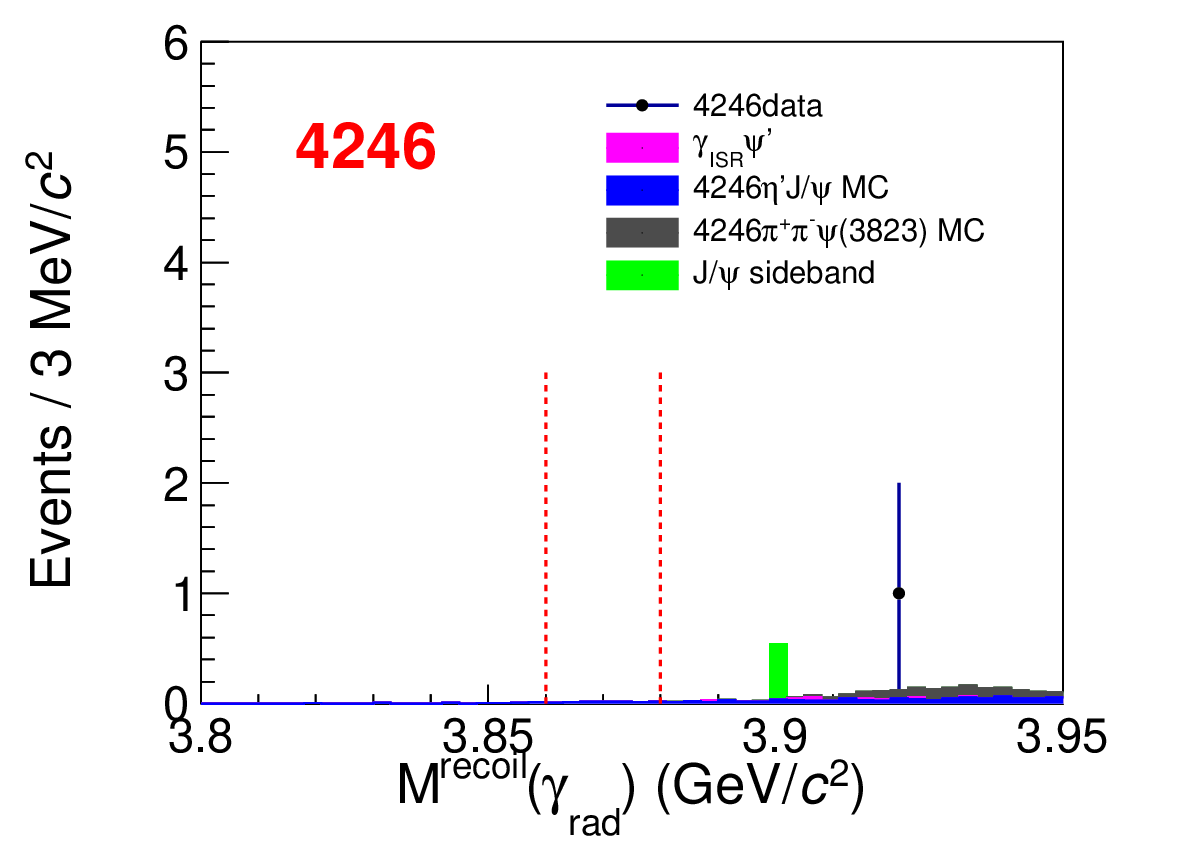}\hspace{5pt}
    \includegraphics[width=0.3\textwidth]{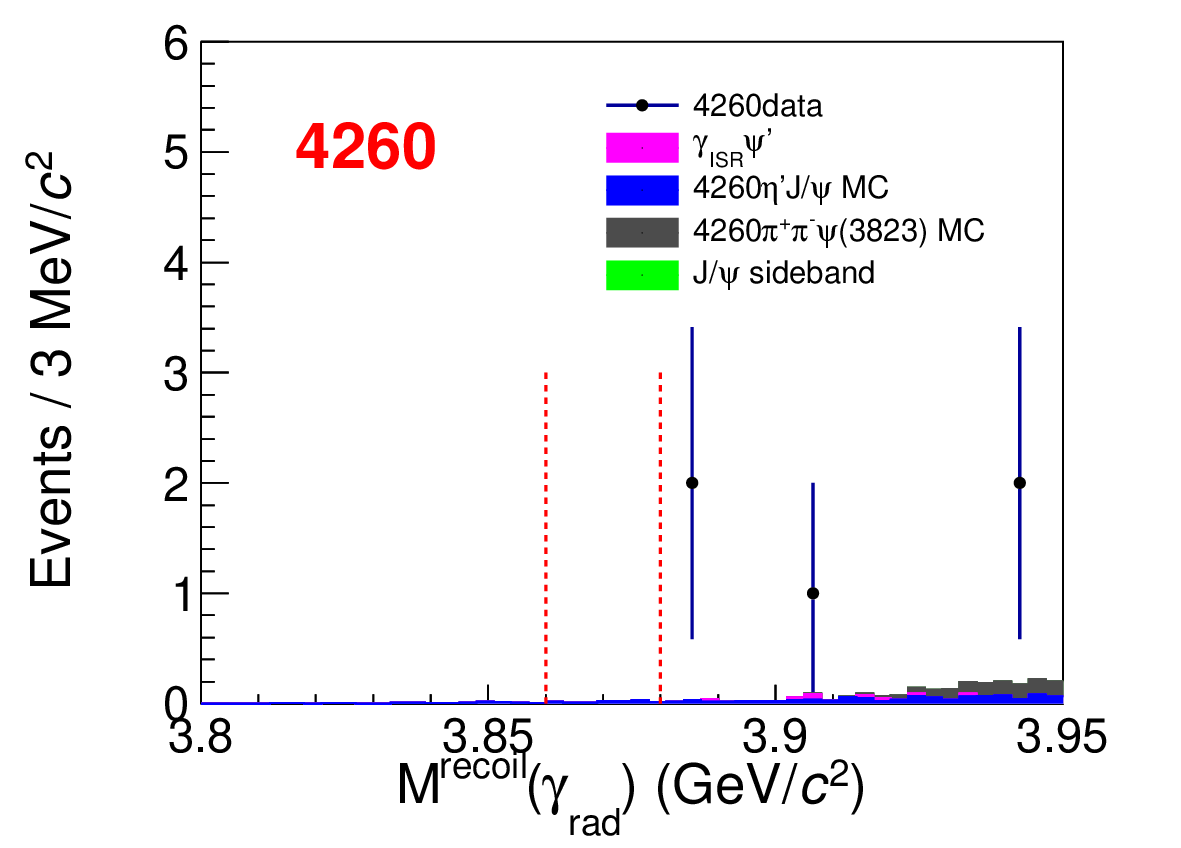}\hspace{5pt}
    \includegraphics[width=0.3\textwidth]{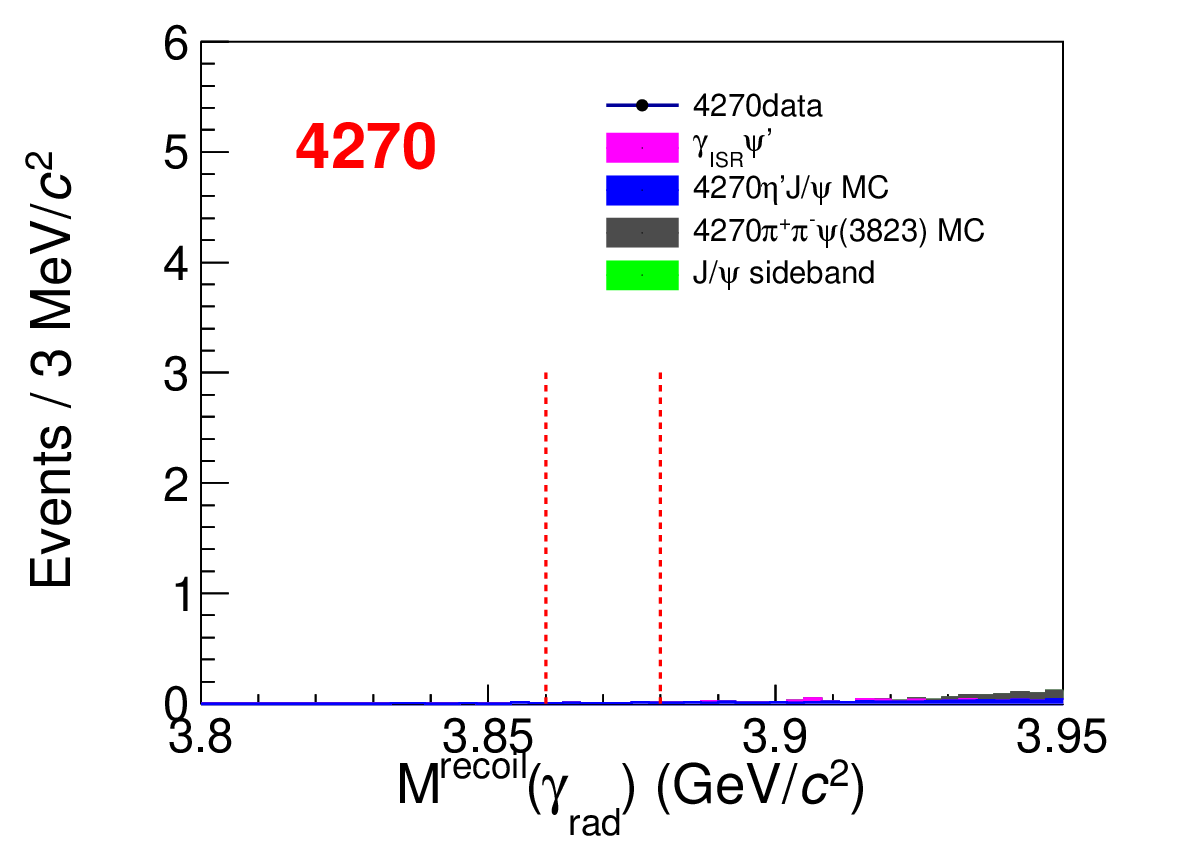}\hspace{5pt}
    \includegraphics[width=0.3\textwidth]{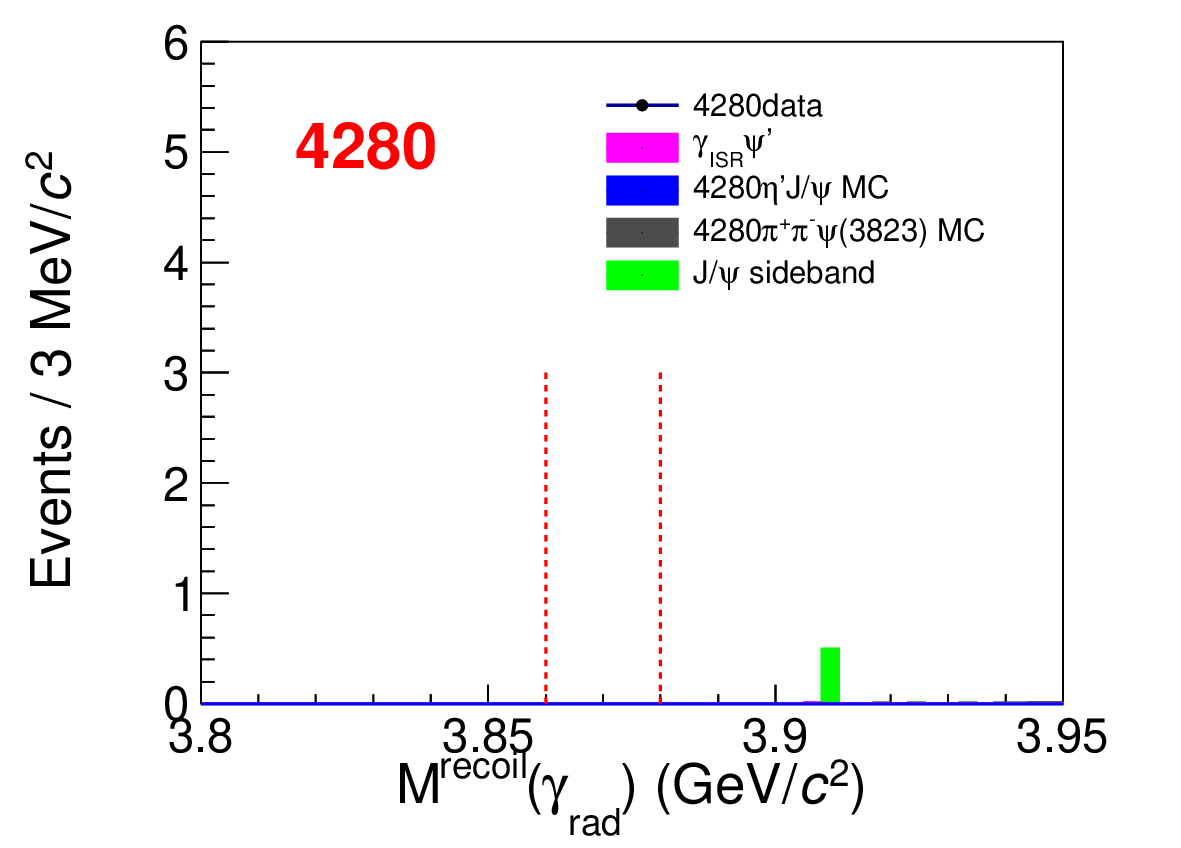}\hspace{5pt}
    \includegraphics[width=0.3\textwidth]{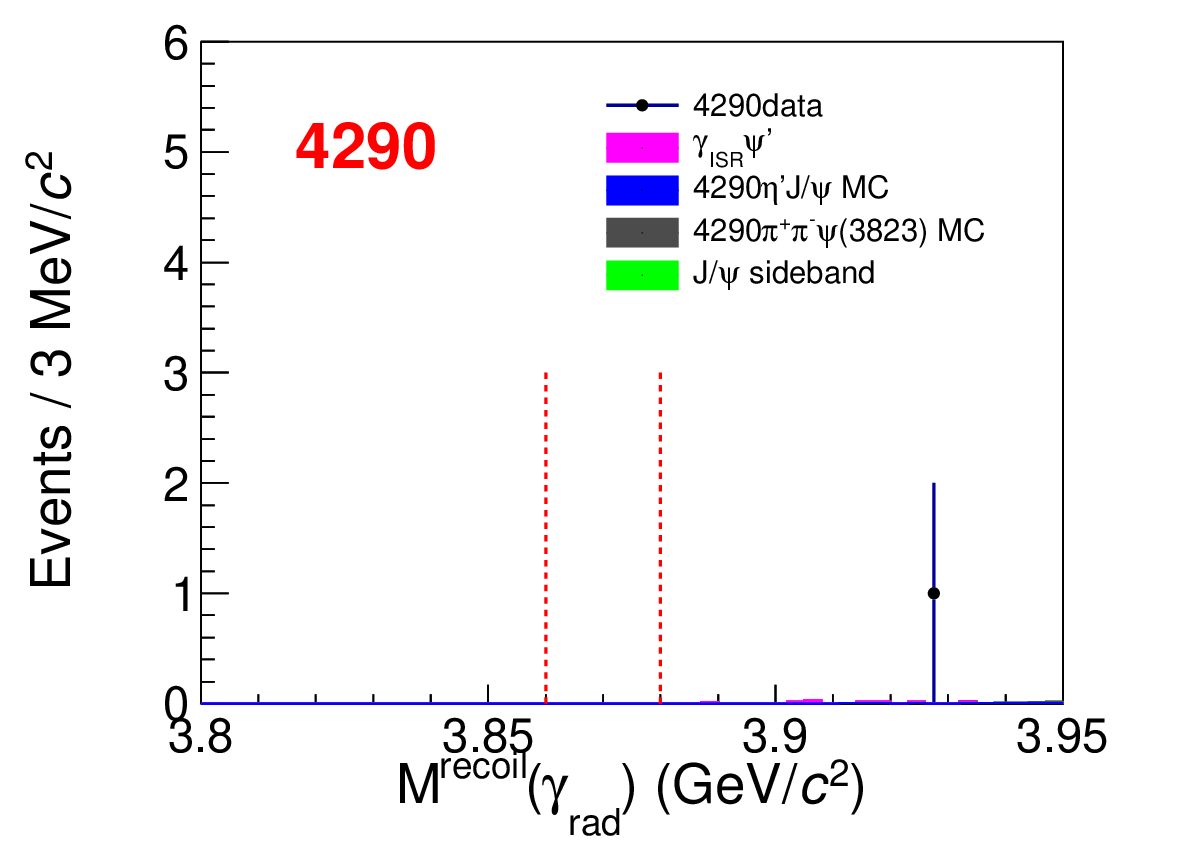}\hspace{5pt}
    \includegraphics[width=0.3\textwidth]{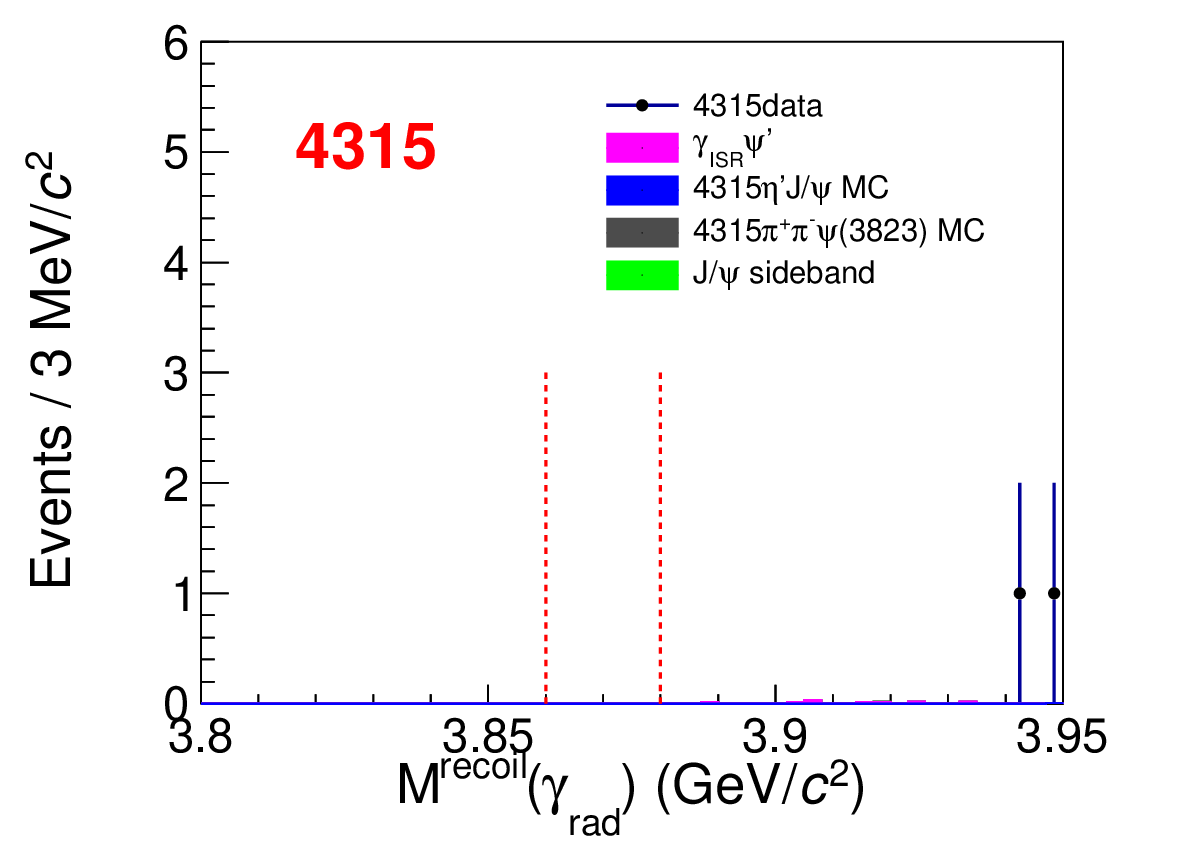}\hspace{5pt}
    \includegraphics[width=0.3\textwidth]{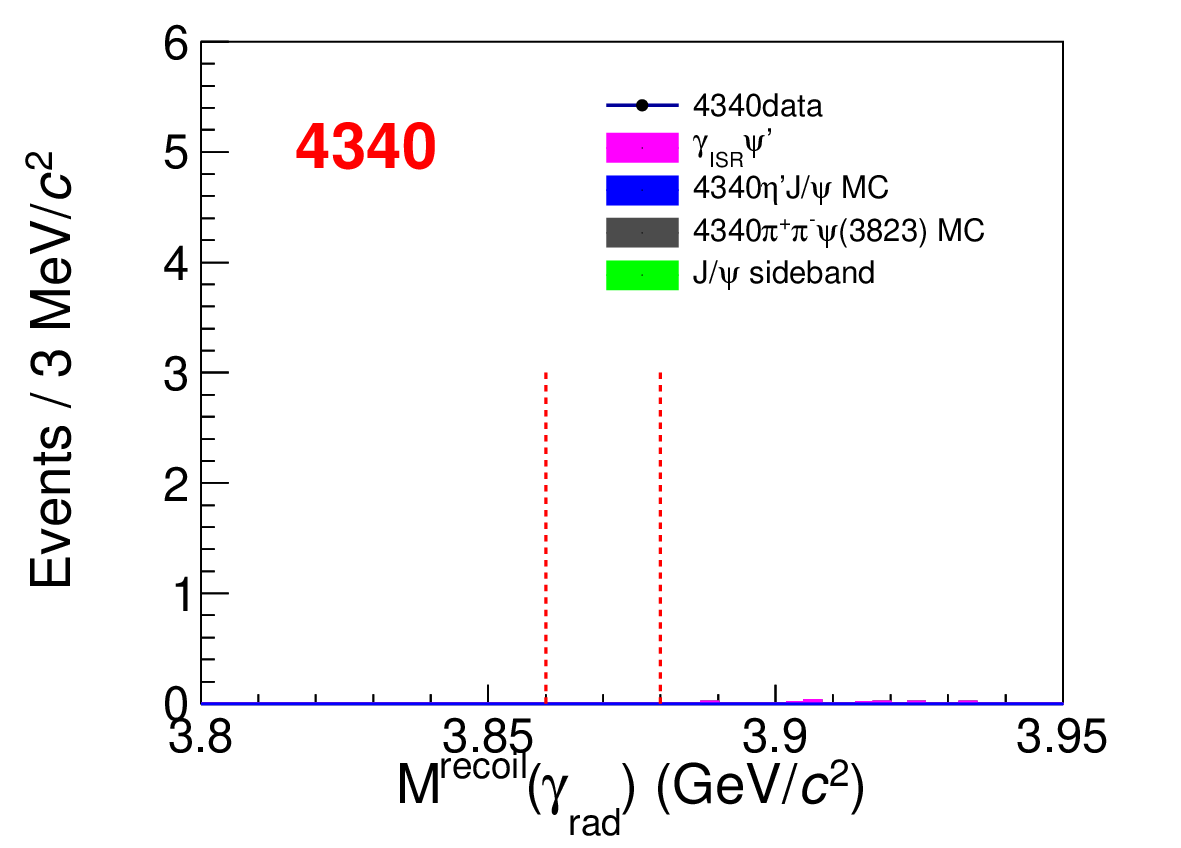}\hspace{5pt}
	\caption{The distribution of $M^{\rm recoil}(\gamma_{\rm rad})$ at each c.m.~energy from 4.16 GeV to 4.34 GeV. For each c.m.~energy, the 3-track events and 4-track events are combined together. Black dots with error bars are data, pink histogram is $\gamma_{\rm ISR}\psi'$ MC, blue histogram is $\eta'\jpsi$ MC, black histogram is $\pp \psi(3823)$ MC and green histogram is $\jpsi$ sideband. The red dotted line represents the signal region of $\x$.}
	\label{rmgy}
\end{figure} 

\section{Systematic uncertainty for cross section measurement}
\label{sys-xs}
\begin{table}[ht]
	\caption{ Systematic uncertainties (in $\%$) for the $\sigma[\ee\to\gamma\x] \, \mathcal{B}[\x\to\pp\chico]$ measurement. The sources marked with "$*$" are shared systematic uncertainties for different data sets.}
	
	\begin{tabular}{c | c | c | c | c | c | c | c | c | c | c | c | c | c | c | c} 
		\hline 
		\hline
		Data set & 4160 & 4180 & 4190 & 4200 & 4210 & 4220 & 4230 & 4237 & 4246 & 4260 & 4270 & 4280 & 4290 & 4315 & 4340\\
		\hline
		Luminosity$^{*}$ & 0.7 & 0.7 & 0.7 & 0.7 & 0.7 & 0.7 & 0.7 & 0.7 & 0.7 & 0.7 & 0.7 & 0.7 & 0.7 & 0.7 & 0.7  \\
		Photon efficiency$^{*}$ & 2.0 & 2.0 & 2.0 & 2.0 & 2.0 & 2.0 & 2.0 & 2.0 & 2.0 & 2.0 & 2.0 & 2.0 & 2.0 & 2.0 & 2.0  \\
		Tracking efficiency$^{*}$ & 2.0 & 2.0 & 2.0 & 2.0 & 2.0 & 2.0 & 2.0 & 2.0 & 2.0 & 2.0 & 2.0 & 2.0 & 2.0 & 2.0 & 2.0  \\
		PID$^{*}$  & 0.5 & 0.5 & 0.5 & 0.5 & 0.5 & 0.5 & 0.5 & 0.5 & 0.5 & 0.5 & 0.5 & 0.5 & 0.5 & 0.5 & 0.5\\
		$\mathcal{B}(\chico\to\gamma\jpsi)^{*}$ & 3.0 & 3.0 & 3.0 & 3.0 & 3.0 & 3.0 & 3.0 & 3.0 & 3.0 & 3.0 & 3.0 & 3.0 & 3.0 & 3.0 & 3.0 \\
		$\mathcal{B}(\jpsi\to\ee/\mm)^{*}$ & 0.6 & 0.6 & 0.6 & 0.6 & 0.6 & 0.6 & 0.6 & 0.6 & 0.6 & 0.6 & 0.6 & 0.6 & 0.6 & 0.6 & 0.6 \\
		$\jpsi$ mass window$^{*}$ & 0.3 & 0.3 & 0.3 & 0.3 & 0.3 & 0.3 & 0.3 & 0.3 & 0.3 & 0.3 & 0.3 & 0.3 & 0.3 & 0.3 & 0.3\\
		Kinematic fit & 1.2 & 1.0 & 1.1 & 1.1 & 1.2 & 1.0 & 1.0 & 1.1 & 1.0 & 1.1 & 1.1 & 1.2 & 1.0 & 1.2 & 1.4\\
		MC decay model$^{*}$ & 2.8 & 2.8 & 2.8 & 2.8 & 2.8 & 2.8 & 2.8 & 2.8 & 2.8 & 2.8 & 2.8 & 2.8 & 2.8 & 2.8 & 2.8\\
		Radiative correction & 0.4 & 1.3 & 1.7 & 1.8 & 1.6 & 1.7 & 1.4 & 1.5 & 1.5 & 1.5 & 1.5 & 1.9 & 1.6 & 1.6 &1.7\\
		\hline
		Total & 5.3 & 5.4 & 5.5 & 5.5 & 5.5 & 5.5 & 5.4 & 5.5 & 5.4 & 5.4 & 5.5 & 5.6 & 5.5 & 5.5 & 5.6\\
		\hline 
		\hline
	\end{tabular}
	\label{systematical error for xs}  
\end{table}	

\end{document}